\documentclass[12pt]{amsart}
\usepackage{amsmath, amssymb}
 \usepackage{epsfig}

\theoremstyle{plain}
\usepackage{amsmath}
\usepackage{amssymb}
\usepackage{amsfonts}
\newtheorem{theorem}{Theorem}

\numberwithin{equation}{section}

\def \< {\langle}
\def \> {\rangle}
\def \^ {\widehat}

\newcommand{\bbW}{{\bf W}}

\newcommand{\bbX}{{\bf X}}

\newcommand{\beq}{\begin{equation}}
\newcommand{\eeq}{\end{equation}}
\newcommand{\bqa}{\begin{eqnarray}}
\newcommand{\eqa}{\end{eqnarray}}
\newcommand{\bqn}{\begin{eqnarray*}}
\newcommand{\eqn}{\end{eqnarray*}}

\newcommand{\bdes}{\begin{description}}
\newcommand{\edes}{\end{description}}
\newcommand{\ep}{\varepsilon}
\def\underwiggle 1{\ifmmode\setbox\TempBox=\hbox{$ 1$}\else\setbox\TempBox=\hbox{1}\fi
\setbox\TempBoxA=\hbox to \wd\TempBox{\hss\char'176\hss}
\rlap{\copy\TempBox}\smash{\lower9pt\hbox{\copy\TempBoxA}} }
\begin{document}

\title[Spectral distributions of Wigner matrices]{New estimators of spectral distributions of Wigner matrices}

\author{Wang Zhou \\
}
\thanks{
  W. Zhou. was partially supported by a grant
R-155-000-106-112 at the National University of Singapore}

\address{Department of Statistics and Applied Probability, National University of
 Singapore, Singapore 117546}
\email{stazw@nus.edu.sg} \subjclass{Primary 15B52, 60F15, 62E20;
Secondary 60F17} \keywords{ Wigner matrices, Stieltjes
transform, nonparametric estimate, domain of attraction of normal law}

\maketitle

\begin{abstract}
We introduce kernel estimators for
the semicircle law.
In this first part of our general theory on the estimators, we prove the consistency
and conduct simulation study to show the performance of the estimators. We also point out that
Wigner's semicircle law for our new estimators and the classical empirical spectral distributions is still true when the elements of Wigner matrices don't have finite variances but are in the domain of
attraction of normal law.

\end{abstract}

\section{Introduction}

\setcounter {equation}{0}
\def\theequation{\thesection.\arabic{equation}}

Random matrix theory gained attention due to work by Wishart in
1928 in mathematical statistics. After that, the subject got
prominence when Wigner began to describe the energy levels of a
 system in quantum mechanics by random matrices. Since then, mathematicians and physicists  have been fascinated by random matrix theory.
Regarding earlier work in this field one may refer to the book by
Mehta \cite{M} and the recent book by Bai and Silverstein \cite{BaiS2010}.

Usually, one assumes that the matrix entries have finite variances. There is little work when the finite second moment condition doesn't hold. We refer to two papers, \cite{BDG} and \cite{BenG}.
This paper is concerned with the Wigner matrix whose units may have infinite variances.

Let $X$ be a complex random variable with $EX=0$. Write
$$l(x)=E|X|^2I(|X|\leq x).$$ Put $b=\inf\{x>0: l(x)>0\}.$
We suppose that $l(x)$ is a slowly varying function as $x\to \infty$.  Define
$$
b_n=\begin{cases} \inf\{x\geq b+1:nl(x)\leq x^2\} \ & \mbox{if}\
E|X|^2=\infty\\
\sqrt{nE|X|^2} &\ \mbox{otherwise}.
\end{cases}
$$
Then $b_n\to \infty$ as $n\to \infty$ if $E|X|^2=\infty$, and $\sum_{j=1}^n X_j/b_n \to_d N(0,1)$ if $X_1,\cdots,X_n$ are independent and identically distributed (i.i.d.) as $X$. (This is the reason why $X$ is said to be in the domain of attraction of normal law.)

The Wigner matrix in this paper is defined
by
\beq \label{wigner}
\bbW_n=b_n^{-1}(X_{ij})_{n\times n},
\eeq
where $X_{ij}=X_{ji}$, and $X_{ij}, 1 \leq i < j < \infty  $ are
i.i.d. complex random variables having the same disttribution as $X$. The diagonal elements $X_{ii}, i=1,2,\cdots$  are i.i.d. real random variables.

The classical limit theorem regarding $\bbW_n$ concerns its
empirical spectral distribution, defined by
$$
F^{\bbW_n}(x)=\frac{1}{p}\sum\limits_{k=1}^pI(\mu_k\leq x),
$$
where $\mu_k,k=1,\cdots,p$ denote the eigenvalues of $\bbW_n$.

Wigner \cite{W55,W58} found the semicircle law as the
limit of $F^{\bbW_n}$ whose distribution $F(x)$ has a density
$$
f(x)=\begin{cases} \frac{1}{2\pi}\sqrt{4-x^2} \ & \mbox{if}\
|x|\leq 2\\
0 &\ \mbox{otherwise}.
\end{cases}
$$

 Clearly we can use $F^{\bbW_n}(x)$ to estimate $F(x)$. However we can't make
any  inference on  $F(x)$ because there
is no central limit theorem about
$\big(F^{\bbW_n}(x)-F(x)\big)$ when the matrix elements only have finite moments. Even the convergence rate of $F^{\bbW_n}$ to $F$ is not clear as far as we know.
This motivates us to find other approaches to understanding the limiting
spectral distribution   $F(x)$.

Another motivation of our current work is the local semicircle law. In \cite{EYY}, the local semicircle law for Wigner matrices, which is a precise estimate for the deviation of the Stieltjes transform of the empirical spectral distribution from the semicircle law,  is obtained under the sub-exponential moment conditions.  One can estimate  the density curve (local semicircle law) based on the observed Wigner matrix by the imaginary part of the Stieltjes transform of the empirical spectral distribution. So is it possible to provide some other estimators for the density curve?

This paper is part of a programme to estimate $F(x)$, as well as $f(x)$ by kernel
estimators. In this paper we will prove the consistency of those
estimators as a first step. We also prove that the semicircle law still holds if we only assume that $E|X|^2I(|X|\leq x)$ is a slowly varying function of $x$ as $x\to \infty$ (which is equivalent to the condition that $X$ is in the domain of attraction of normal law).

\section{Methodology and Main Results}

Suppose that the observations $X_1,\cdots,X_n$ are i.i.d random
variables with an unknown density function $g(x)$ and $F_n(x)$ is
the empirical distribution function determined by the sample.
Then, a popular nonparametric estimate of $g(x)$ is
\begin{equation}\label{a24}
\hat{f}_n(x)=\frac{1}{nh}\sum\limits_{j=1}^nK(\frac{x-X_j}{h})=\frac{1}{h}\int
K(\frac{x-y}{h})dF_n(y),
\end{equation}
where the function $K(y)$ is a Borel function and $h=h(n)$ is the
bandwidth which tends to $0$ as $n\to \infty$. Obviously,
$\hat{f}_n(x)$ is again a probability density function and
moreover, it inherits some smooth properties of $K(x)$, provided that
the kernel is taken as a probability density function. Under some
regular conditions on the kernel, it is well-known that
$\hat{f}_n(x)\rightarrow g(x)$ in some sense (with probability one
or in probability). There is a huge literature regarding this kind
of estimate. For example one may refer to Rosenblatt \cite{R1},
Parzen \cite{p1}, Hall \cite{peter} and the book by Silverman
\cite{bw}.

Enlightened by (\ref{a24}), we propose the following estimator
$f_n(x)$ of $f(x)$, given by
\begin{eqnarray}
f_n(x)&=&\frac{1}{ph}\sum\limits_{i=1}^pK(\frac{x-\mu_i}{h})=\frac{1}{h}\int
K(\frac{x-y}{h})dF^{\bbW_n}(y)\label{a1}, \label{fnx}\\
F_n(x)&=&\int^x_{-\infty}\,f_n(t)\,dt,  \label{Fnx}
\end{eqnarray}
where $\mu_i,i=1,\cdots,p$ are eigenvalues of $\bbW_n$.
 It turns out that
$f_n(x)$ and $F_n(x)$ are respective consistent estimators of
 $f(x)$ and $F(x)$ under some regular conditions. It is interesting to note that if
 \begin{equation} \label{cauchy}
 K(x)=\pi^{-1}(1+x^2)^{-1},
 \end{equation}
 then
 $$
 f_n(x)=\pi^{-1}\Im \int (\lambda-x-ih)^{-1} d\,F^{\bbW_n}(\lambda).
 $$
 So the imaginary part of the Stieltjes transform is a special case of our kernel estimators.

Before we present our main results, we introduce some regular conditions on the kernel function. Throughout the paper, suppose that the kernel function $K(x)$ satisfies
\begin{equation}
\label{a26} K(x)\geq 0, \ \ \int K(x)dx=1,
\end{equation}
\begin{equation}
\label{a27}\int |K'(x)|\,dx<\infty.
\end{equation}

\begin{theorem} \label{t1}
Let $\bbW_n$ be defined by \eqref{wigner}. Suppose that $K(x)$ satisfies
 (\ref{a26}).  Let $h=h(n)$ be a sequence of
positive constants tending to zero. Then,
\beq \label{con-dis}
\|F_n-F\|:=\sup_x |F_n(x)-F(x)| \to 0
\eeq
almost surely as $n\to \infty$.
\end{theorem}

Theorem \ref{t1} follows from the following result.

\begin{theorem} \label{t2}
Let $\bbW_n$ be defined by \eqref{wigner}. Then $F^{\bbW_n}$ tends to the semi-circlular law almost surely.
\end{theorem}

In the literature, Theorem \ref{t2} was derived under the assumption that $E|X|^2<\infty$. See \cite{BaiS2010}. In \cite{BenG}, Ben Arous and Guiomet considered the spectrum of heavy tailed random matrices. The proved that the spectral distribution of $\bbX_n=(X_{jk})$ normalized by a suitable constant tends to a limit $\mu_{\alpha}$ when $X$ is in the domain of attraction of a $\alpha$ stable law ($0<\alpha<2$). $\mu_{\alpha}$ is also characterized as having infinite support and being absolute continuous with respect to the Lebesgue measure. However from our Theorem \ref{t2}, we see that $\mu_2$ is still the semicircle law. This is consistent with the result that $\mu_\alpha$ converges to the semicircle law \cite{BDG} as $\alpha$ tends to $2$ from below.

The final result is about the convergence of the density estimator. Unlike Theorems \ref{t1} or \ref{t2}, we need more moment conditions in order to show the convergence.

\begin{theorem} \label{theo3} Let $\bbW_n$ be defined by \eqref{wigner}. Suppose that $K(x)$ satisfies
 \eqref{a26} and \eqref{a27}.  Let $h=h(n)$ be a sequence of
positive constants tending zero such that $nh^2\to \infty$.
Assume that $E|X_{11}|^3<\infty, E|X_{12}|^6<\infty$.
Then
$$\sup\limits_{x\in [-2,+2]}|f_n(x)- f(x)| \longrightarrow 0 \ \ \text{in probability}.$$
\end{theorem}

\section{Simulation Study}
In this section, we perform a simulation study to investigate the behavior of the kernel estimators of  the semi-circle law.

We consider two different populations, exponential and Poisson distributions. From each population, we generate two samples of size $50\times 51/2$ and $800\times 801/2$ respectively. So we can form two symmetric random matrices $(X_{ij})_{50\times 50}$ and $(X_{ij})_{800\times 800}$.
The kernel is selected as
$$
K(x)=(2\pi)^{-1/2}e^{-x^2/2},
$$
which is the standard normal density function. Here we don't choose the kernel as \eqref{cauchy} since it is well known that the imaginary part of the Stieltjes transform divided by $\pi$ tends to the density function. The bandwidth is chosen as $h=n^{-2/5}$ ($n=50, 800$).

For $(X_{ij})_{50\times 50}$, the kernel density estimator is
$$
\frac1{50\times 50^{-2/5}}\sum_{i=1}^{50} K\big((x-\mu_i)/50^{-2/5}\big),
$$
where $\mu_i, i=1,\cdots,50$ are eigenvalues of $50^{-1/2}(X_{ij})_{50\times 50}$. This curve is drawn by dotted lines in Figures 1-2.

For $(X_{ij})_{800\times 800}$, the kernel density estimator is
$$
\frac1{800\times 800^{-2/5}}\sum_{i=1}^{800} K\big((x-\mu_i)/800^{-2/5}\big),
$$
where $\mu_i, i=1,\cdots,800$ are eigenvalues of $800^{-1/2}(X_{ij})_{800\times 800}$. This curve is drawn by dashed lines in Figures 1-2.

The density function of the semicircle law is drawn by solid lines in Figures 1-2. Here in Figure 1, the distribution of $X_{11}$ is
\begin{equation} \label{dist}
P(X_{11}\leq x)=1-e^{-(x+1)}, \ \ x\geq -1.
\end{equation}
In Figure 2, the distribution is
$$
P(X_{11}=k)=e^{-1}/(k+1)!, \ \ k=-1,0,1,\cdots.
$$
From the two pictures, we see that  the estimated curves fit the semicircle law very well.
As $n$ becomes large, the estimated curves become closer to the semicircle law.

In Figure 3, for the exponential distribution,
we compare the kernel distribution estimator
$$
\int_{-\infty}^y \frac1{50\times 200^{-2/5}}\sum_{i=1}^{50} K\big((x-\mu_i)/50^{-2/5}\big)dx,
$$
which is plotted by dotted line, the empirical spectral distribution function $50^{-1}\sum_{i=1}^{50}I(\mu_i \leq x)$ which is plotted by
the dashed line, and the semicircle law distribution function by the solid line. We see that although the two estimated curves are very close to the semicircle law, the kernel distribution estimator performs better.

In Figure 4, based on the eigenvalues $\mu_1, \cdots,\mu_{50}$ of $W_{50}$, we compare the kernel distribution estimator (dotted line) $F_{50}(y)=\int_{-\infty}^y\, \frac1{50\times 50^{-2/5}}\sum_{i=1}^{50}$ $K\big((x-\mu_i)/50^{-2/5}\big)dx$, the empirical spectral distribution (dashed line) $ F^{\bbW_{50}}(y)=50^{-1}\sum_{i=1}^{50}I(\mu_i \leq y)$, and the semi-circle law distribution function. Generally speaking, $F_{50}$ is closer to the semi-circle law than $ F^{\bbW_{50}}$.

\vspace{1in}

\centerline{(Place Figures 1--4 here.)}

\vspace{1in}

\section{Proof of Theorems}
\noindent
{\bf Proof of Theorem \ref{t1}}.
The Stieltjes transform of $F_n(x)$ is
\begin{eqnarray}
\int\, (x-z)^{-1}f_n(x)\,dx&=&(ph)^{-1}\sum_{j=1}^p \int\,(x-z)^{-1}K((x-\mu_j)/h)\,dx \nonumber\\
&=&\int\! \int\, (\lambda+yh-z)^{-1}K(y)\,dy\,dF^{\bbW_n}(\lambda). \label{ch1}
\end{eqnarray}
where $t\in R$. For arbitrary $\ep>0$, we choose $y_0$ such that
\[
\int_{|y|>y_0}K(y)\,dy \leq \ep.
\]
So
\beq \label{out}
\int\! \int_{|y|\geq y_0}\, |(\lambda+yh-z)^{-1}|K(y)\,dy\,dF^{\bbW_n}(\lambda) \leq \ep/\Im z.
\eeq
Also note that
\begin{eqnarray}
&&\int\! \int_{|y|\leq y_0}\, |(\lambda+yh-z)^{-1}-(\lambda-z)^{-1}|K(y)\,dy\,dF^{\bbW_n}\nonumber\\
&=&\int\!\int_{|y|\leq y_0} \frac{|yhK(y)|}{|\lambda-z||\lambda+yh-z|}\,dy\,dF^{\bbW_n}(\lambda)\nonumber\\
&\leq&hy_0/(\Im z)^2. \label{in}
\end{eqnarray}
Combining \eqref{ch1}-\eqref{in}, we have
\begin{eqnarray*}
\lim_{n\to \infty} \Big|\int\, (x-z)^{-1}f_n(x)\,dx-\int\,(\lambda-z)^{-1}\,dF^{\bbW_n}(\lambda)\Big| \leq 2\ep/\Im z.
\end{eqnarray*}
Since $\ep$ is arbitrary, we can complete the proof. \qed

\

\noindent
{\bf Proof of Theorem \ref{t2}}. The finite variance case is already treated in the literature. We only consider the case where $E|X|^2=\infty$.
If $l(x)=EX^2I(|X|\leq x)$ is slowly varying as $x\to \infty$, then
\begin{equation} \label{fact1}
P(|X|\geq x)=o(l(x)/x^2), \ \ E|X|I(|X|\geq x)=o(l(x)/x)
\end{equation}
as $x\to \infty$.

We prove the theorem by the following steps.

{\bf Step 1. Removing the diagonal elements}

Let $\tilde \bbW_n$ be the matrix derived from $\bbW_n$ by replacing the diagonal elements with zero. We will show that the limiting spectral distributions of the two matrices are the same if one of them exists.

Let $N_n=\sum_{k=1}^n I(|X_{kk}| \geq b_n^{1/2})$. Replace the diagonal elements of $\bbW_n$ by $b_n^{-1}x_{kk}I(|X_{kk}| \leq b_n^{1/2})$ and denote the resulting matrix by $\hat \bbW_n$. Then the L\'evy distance $L(F^{\tilde \bbW_n},F^{\hat \bbW_n})$ satisfies
$$
L^3(F^{\tilde \bbW_n},F^{\hat \bbW_n})\leq n^{-1} \text{tr}[(\tilde \bbW_n-\hat \bbW_n)^2]\leq (nb_n^2)^{-1}\sum_{k=1}^n x_{kk}^2I(|X_{kk}| \leq b_n^{1/2}) \leq b_n^{-1}.
$$
The rank inequality gives
$$
\|F^{\tilde \bbW_n},F^{ \bbW_n}\| \leq N_n/n.
$$
So it suffices to show that $N_n/n \to 0 $ almost surely. Write $p_n=P(|X_{11}| \geq b_n^{1/2})\to 0$. From Bernstein inequality, for any $\ep>0$, we have
\begin{eqnarray*}
P(N_n\geq \ep n)\leq 2\exp\Big(-(\ep-p_n)^2n^2/[2np_n+2(\ep-p_n)n]\Big)\leq e^{-bn}
\end{eqnarray*}
for some positive constant $b$. This completes the first-step proof.

In the following, we can assume that the diagonal elements of $\bbW_n$ are zero.

\

{\bf Step 2. Truncation}

Let $\widehat \bbW_n=b_n^{-1}(x_{jk}I(|x_{jk}|\leq  b_n))$. By the rank inequality,
\begin{eqnarray}
\|F^{\bbW_n}-F^{\widehat \bbW_n}\| &\leq & n^{-1}\text{rank}(\bbW_n-\widehat \bbW_n) \nonumber\\
&\leq & 2n^{-1}\sum_{1\leq j\leq k\leq n}I(|x_{jk}|\geq  b_n). \label{trun1}
\end{eqnarray}
By \eqref{fact1}, we have
$$
E\Big(2n^{-1}\sum_{1\leq j\leq k\leq n}I(|x_{jk}|\geq  b_n)\Big) =o(nl(b_n)/b_n^2)=o(1),
$$
and
$$
\text{Var}\Big(2n^{-1}\sum_{1\leq j\leq k\leq n}I(|x_{jk}|\geq  b_n)\Big)\leq 2P(|X_{12}|>b_n)=o(n^{-1}).
$$
Then , applying Berstein inequality, for all small $\ep>0$ and large $n$, we have
\beq \label{trun2}
P\Big(n^{-1}\sum_{1\leq j\leq k\leq n}I(|x_{jk}|\geq  b_n) \geq \ep \Big) \leq 2e^{-\ep n},
\eeq
which is summable. By \eqref{trun1} and \eqref{trun2}, it suffices to show that $F^{\widehat \bbW_n}$ converges to the semicircle law.

\

{\bf Step 3. Normalization}
We consider the L\'evy distance between $F^{\widehat \bbW_n}$ and $F^{\widehat \bbW_n-E\widehat \bbW_n}$ which tends to zero since
\begin{eqnarray*}
&&L^3(F^{\widehat \bbW_n},F^{\widehat \bbW_n-E\widehat \bbW_n}) \nonumber \\
&\leq& (nb_n^2)^{-1}\sum_{j\not=k}  \big(Ex_{jk}I(|X_{jk}| \leq b_n) \big)^2 \nonumber \\
&\leq&1/b_n^2 \to 0.
\end{eqnarray*}
Also noting that
$$\text{Var}\big(X_{12}I(|X_{12}|\leq b_n)\big)=l(b_n)-\big(EX_{12}I(|X_{12}|\leq b_n)\big)^2=l(b_n)\big(1+o(n^{-1})\big)$$
by \eqref{fact1}, we conclude that
$$
\|F^{\widehat \bbW_n-E\widehat \bbW_n}-F^{\overline \bbW_n}\| \to 0
$$
as $n\to \infty$, where
$$
\overline \bbW_n=\Big(n\text{Var}\big(X_{12}I(|X_{12}|\leq b_n)\big)\Big)^{-1/2}\Big( x_{jk}I(|x_{jk}|\leq  b_n)-Ex_{jk}I(|x_{jk}|\leq  b_n)\Big).
$$

{\bf Step 4. Complete the proof}. From the above three steps, we know that it remains to prove that $F^{\overline \bbW_n}$ tends to the semicircle law.  Let
$$
Y_{jk}=[\text{Var}\big(X_{12}I(|X_{12}|\leq b_n)\big)]^{-1/2}[ x_{jk}I(|x_{jk}|\leq  b_n)-Ex_{jk}I(|x_{jk}|\leq  b_n)].
$$
Then for arbitrary $\eta>0$, by \eqref{fact1} and $l(\eta b_n)/l(b_n)\to 1$,
$$
E|Y_{12}|^2I(|Y_{12}|>\eta \sqrt n) \to 0
$$
as $n\to\infty$. This implies that
$$
\lim_{n\to \infty} n^{-2} \sum_{jk}E|Y_{jk}|^2I(|Y_{jk}|>\eta \sqrt n) \to 0
$$
as $n\to\infty$.  So we can conclude the proof by Theorem 2.9 in \cite{BaiS2010}.
\qed

\

\noindent
{\bf Proof of Theorem \ref{theo3}}.
Using integration by parts we obtain
\begin{eqnarray*}
&&|\frac{1}{h}\int K(\frac{x-t}{h})dF^{\bbW_n}(t)-\frac{1}{h}\int
K(\frac{x-t}{h})dF(t)|\\
&\leq&|\frac{1}{h^2}\int
K'(\frac{x-t}{h})\Big(F^{\bbW_n}(t)-F(t)\Big)dt| \\
&=&O_p\Big(\frac{1}{\sqrt{n}h}\Big),\\
\end{eqnarray*}
as $n\to \infty$, where the last step uses Theorem 1.1 in \cite{BHPZ} and \eqref{a27}. Finally,
$$
 |\frac{1}{h}\int K(\frac{x-t}{h})dF(t)-f(x)|
$$
$$
=|\int \Big(f(x-t)-f(x)\Big)\frac{1}{h}K(\frac{t}{h})dt|
$$
$$
\leq \sup\limits_{x\in [-2,2]}|\int_{|t|>\delta} \Big(f(x-t)-f(x)\Big)\frac{1}{h}K(\frac{t}{h})dt|
$$
$$
+\sup\limits_{x\in [-2,2]}|\int_{|t|\leq\delta}
\Big(f(x-t)-f(x)\Big)\frac{1}{h}K(\frac{t}{h})dt|
$$
$$
\leq 2\sup\limits_{x\in [-2,2]}f(x)\int_{|y|>
\delta/h}|K(y)|dy+\sup\limits_{x\in
[-2,2]}\sup\limits_{|t|\leq\delta}|(f(x-t)-f(x)|\int\frac{1}{h}|K(\frac{t}{h})|dt,
$$
which goes to zero by noting that $f(x)$ is continuous,
choosing sufficiently small $\delta$ and then for such $\delta$
letting $n\rightarrow\infty$.  Thus the proof is complete. \qed

\newpage

 \begin{figure}
 \centerline{Figure 1: Spectral density curves for Wigner matrices }
 \centerline{$n^{-1/2}(X_{ij})_{n\times n}$, $X_{ij}\sim$ exponential distribution }
\includegraphics[height=12cm,width=14cm]{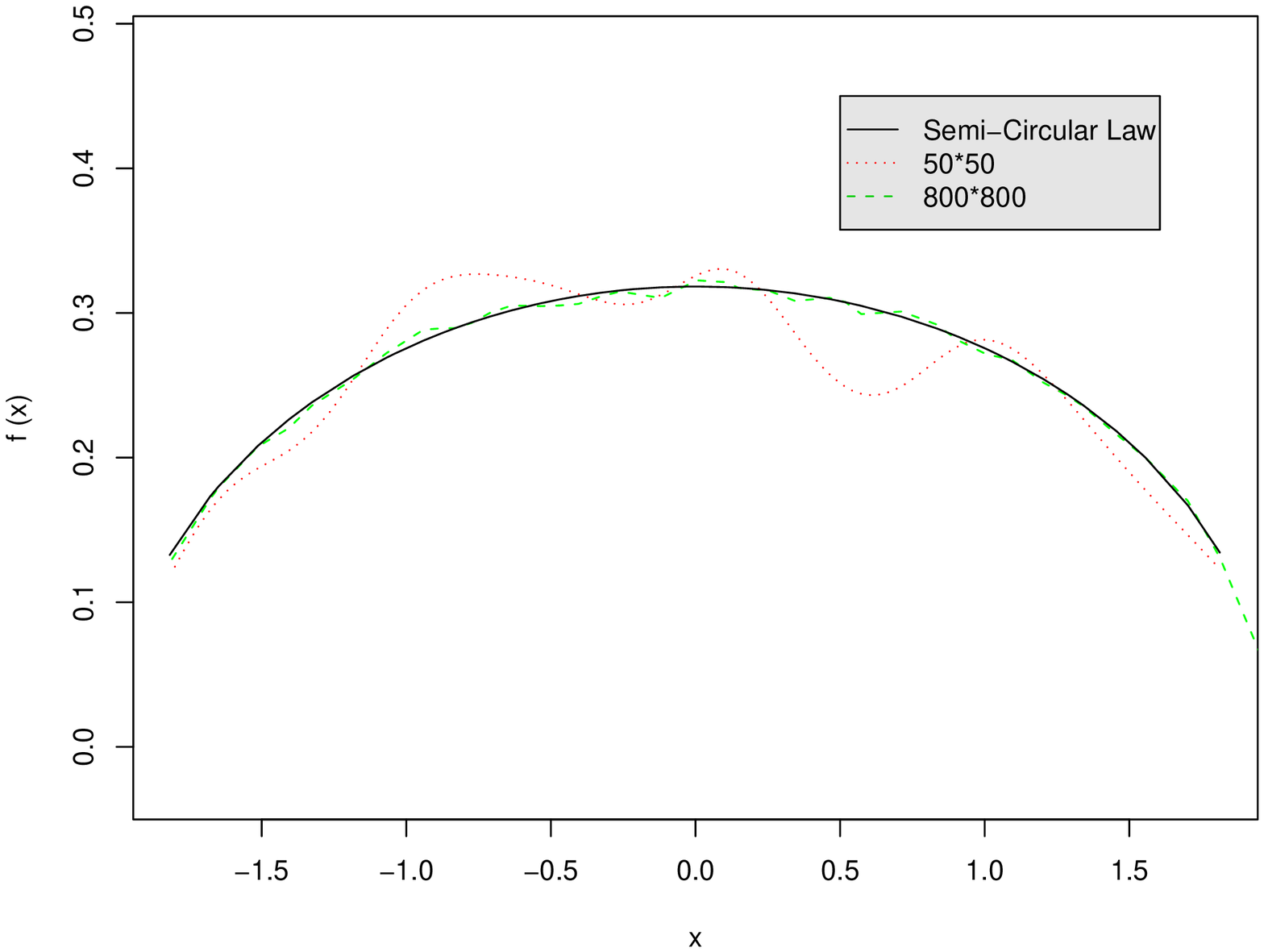}
 \vspace{7cm}
\end{figure}

\newpage
\begin{figure}
 \centerline{Figure 2: Spectral density curves for Wigner matrices}
 \centerline{$n^{-1/2}(X_{ij})_{n\times n}$, $X_{ij}\sim$ Poisson distribution }
\includegraphics[height=12cm,width=14cm]{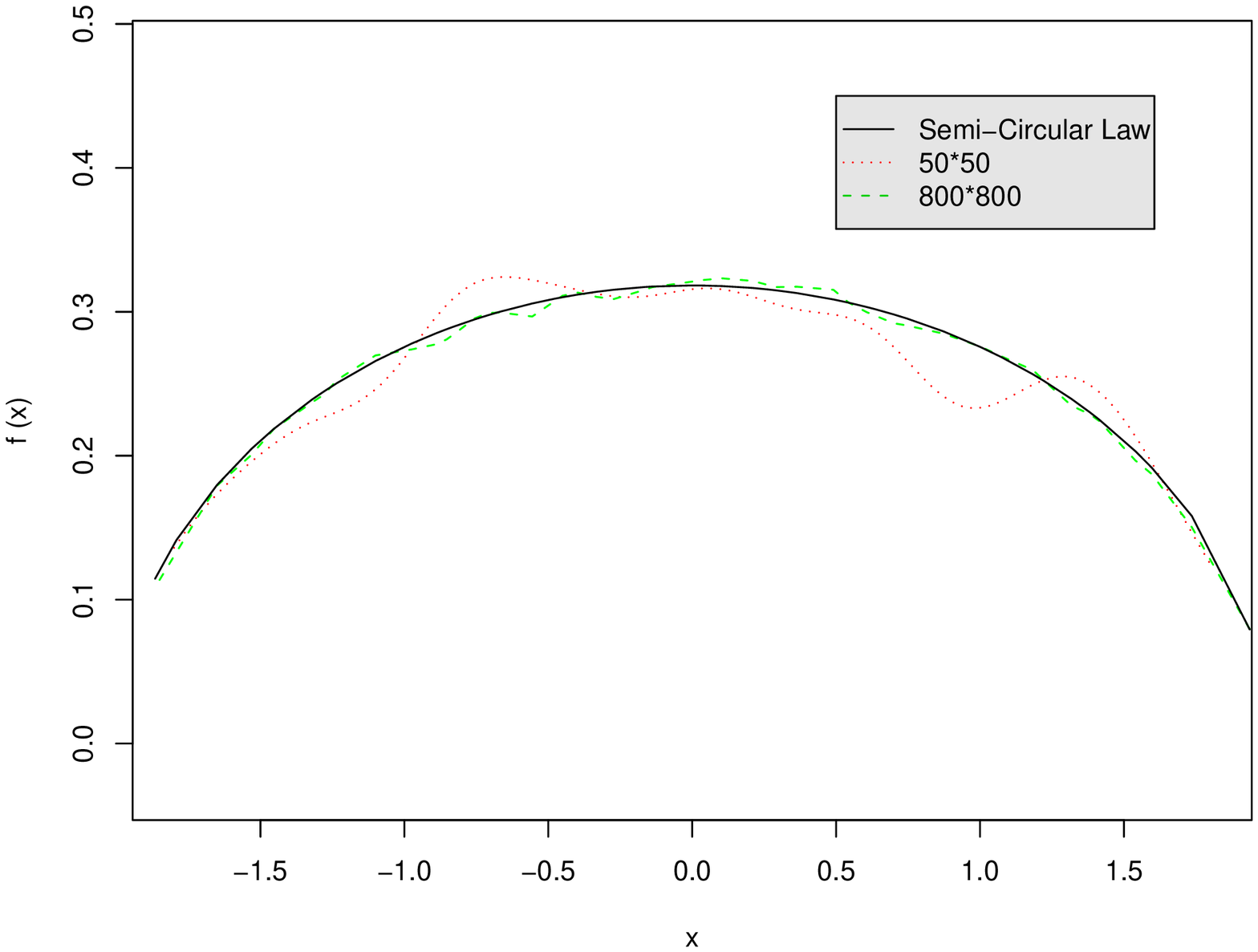}
 \vspace{7cm}
\end{figure}

\newpage
\begin{figure}
\centerline{Figure 3: Distribution curves for Wigner matrices}
 \centerline{$50^{-1/2}(X_{ij})_{50\times 50}$, $X_{ij}\sim$ exponential distribution }
\includegraphics[height=12cm,width=14cm]{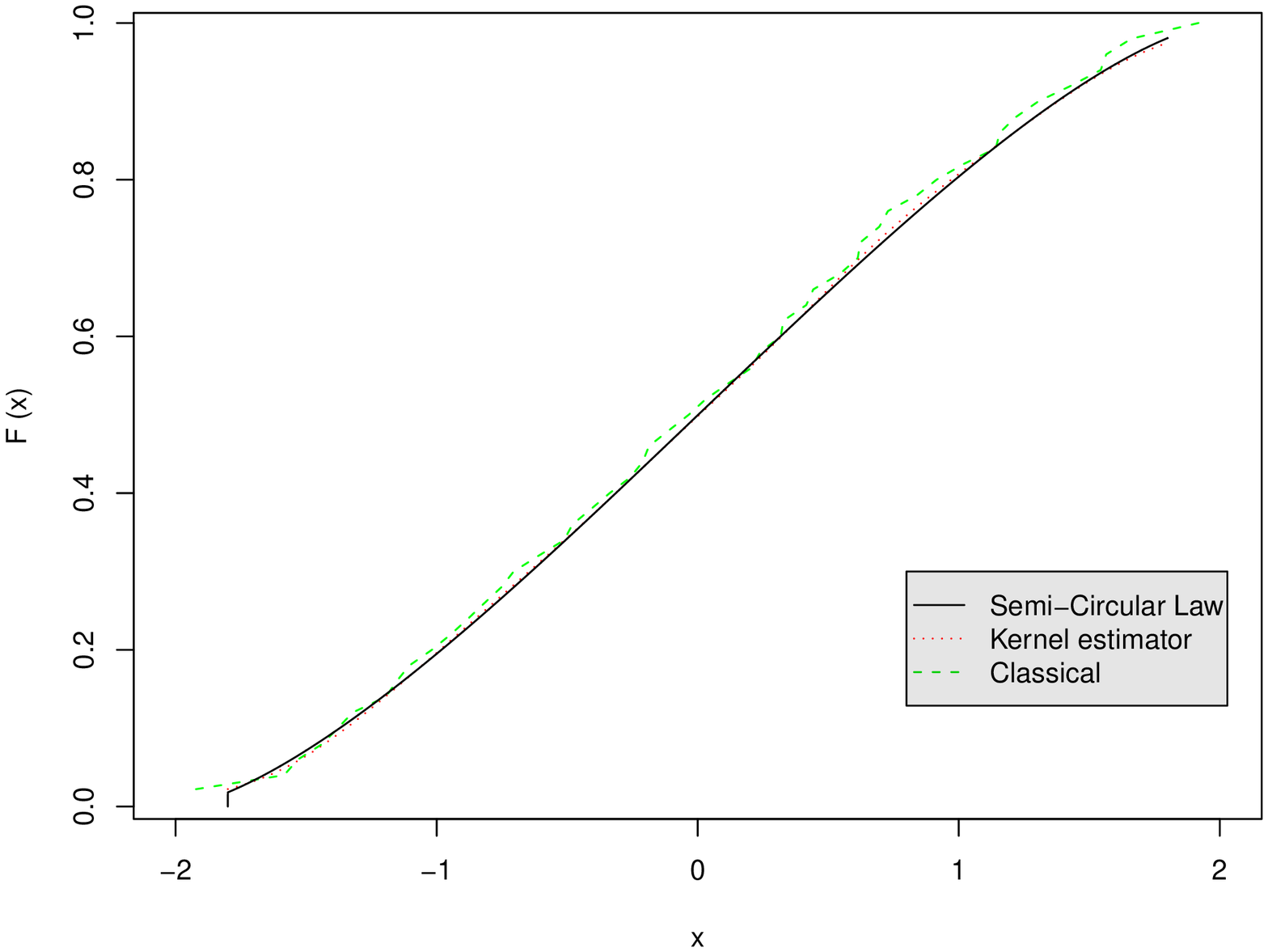}
\vspace{7cm}
\end{figure}

\newpage
 \begin{figure}
\centerline{Figure 4: Distribution curves for Wigner matrices}
\centerline{$50^{-1/2}(X_{ij})_{50\times 50}$, $X_{ij}\sim$ Poisson distribution}
\includegraphics[height=12cm,width=14cm]{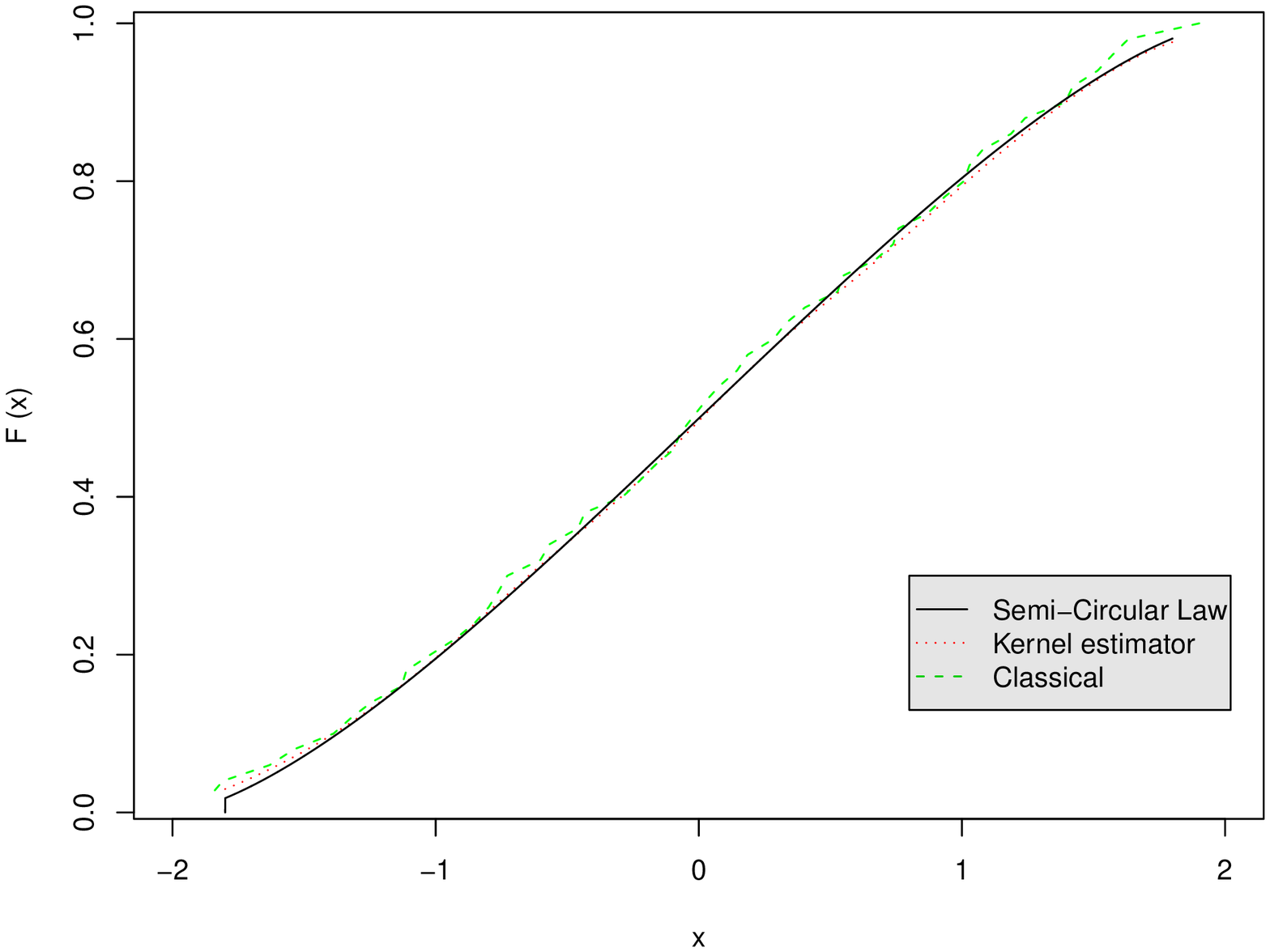}
\vspace{7cm}
\end{figure}

\end{document}